\documentclass[twocolumn]{aastex631}

\usepackage{natbib}
\usepackage{epsfig}
\usepackage{graphicx}
\usepackage{subfigure}
\usepackage{float}
\usepackage{amsmath}
\usepackage{color}
\usepackage{amssymb}
\usepackage{apjfonts}
\usepackage{enumerate}

\newcommand{\PMO}{Purple Mountain Observatory, Chinese Academy of Sciences, Nanjing 210023, China}
\newcommand{\USTC}{School of Astronomy and Space Sciences, University of Science and Technology of China, Hefei 230026, China}
\DeclareRobustCommand{\forceLower}[1]{\MakeLowercase{#1}}

\shortauthors{Wang \& Wei}

\graphicspath{{./}{figures/}}

\usepackage{amsmath}
\begin{document}

\title{The Dispersion Relation of Massive Photons in Plasma: \\
	A Comment on ``Bounding the Photon Mass with Ultrawide Bandwidth Pulsar Timing Data and
	Dedispersed Pulses of Fast Radio Bursts''}

\correspondingauthor{Jun-Jie Wei}
\email{jjwei@pmo.ac.cn}

\author[0000-0003-3635-5375]{Bao Wang}
\affiliation{\PMO}
\affiliation{\USTC}

\author[0000-0003-0162-2488]{Jun-Jie Wei}
\affiliation{\PMO}
\affiliation{\USTC}

\begin{abstract}

The dispersion measures of fast radio bursts have been identified as a powerful tool for testing the zero-mass hypothesis 
of the photon. The classical approach treats the massive photon-induced and plasma-induced time delays as two separate phenomena.
Recently, \cite{Wang2024} suggested that the joint influence of the nonzero photon mass and plasma effects should be considered, and proposed a revised time delay for massive photons propagating in a plasma medium, denoted as $\Delta t'_{m_{\gamma}} \propto \nu^{-4}$, which departures from the classical dispersion relation ($\propto \nu^{-2}$).
Here we discuss the derivation presented by \cite{Wang2024} and show that the classical dispersion relation 
remains valid based on Proca equations.

\end{abstract}


\section{INTRODUCTION}
\label{Sec:Intro}

Since the discovery of fast radio bursts (FRBs), numerous studies have employed the dispersion measures (or equivalently, the frequency-dependent time delays) of FRBs to place upper limits on the rest mass of the photon ($m_{\gamma}$; \citealt{Wu2016, Bonetti2016, Bonetti2017, Shao2017, Xing2019, Wei2020, Wang2021, Lin2023, Wang2023, Ran2024}).
In the classical treatment, two separate time delays are considered: the plasma-induced time delay ($\Delta t_{\rm d} \propto \nu^{-2}$) and the massive photon-induced time delay ($\Delta t_{m_{\gamma}} \propto \nu^{-2}$).
These two contributions are considered to be directly additive.
However, in a recent work, \cite{Wang2024} proposed a revised time delay, denoted as $\Delta t_{m_{\gamma}}' \propto \nu^{-4}$, which is attributed to the joint influence of the nonzero photon mass and plasma effects.
In this comment, we point out the contradictions in the revised time delay proposed by \cite{Wang2024}, and derive the dispersion relation of massive photons propagating in plasma based on Proca equations \citep{Proca1936}, thereby proving the validity of the classical time delay ($\propto \nu^{-2}$).

~\\
\section{Comment on Time Delay $\forceLower{\Delta t_{m_{\gamma}}'} \propto \nu^{-4} $}
\label{Sec:comment}

In this section, we follow the derivation presented in \cite{Wang2024} to examine the revised time delay.
We first define four types of arrival times:
\begin{enumerate}[(i)]
	\item The arrival times of massless photons in vacuum and plasma are $t_{1}$ and $t_{2}$, respectively.
	\item  The arrival times of massive photons in vacuum and plasma are $t_{1}' =\int \frac{1}{\upsilon_{\rm g}^{\gamma}} {\rm d} l$ and $t_{2}' =\int \frac{1}{\upsilon_{\rm g}^{\rm new}} {\rm d} l$, respectively, where $\upsilon_{\rm g}^{\gamma}$ and $\upsilon_{\rm g}^{\rm new}$ are the group velocities of massive photons in vacuum and plasma, respectively.
\end{enumerate}
According to \cite{Wang2024}, the observed time delay between massive photons with different frequencies ($\nu_{\rm low}$, $\nu_{\rm high}$; $\nu_{\rm low}<\nu_{\rm high}$) in a plasma medium is expressed as
\begin{eqnarray}\label{new relation}
		\Delta t_{\mathrm{obs}}^{\prime} 
		&=& \left(t_{\rm 2, low}'-t_{\rm 1, low}' \right) - \left( t_{\rm 2, high}'-t_{\rm 1, high}' \right) \nonumber \\
		&=&  \int \left(\frac{1}{\upsilon_{\mathrm{g}, \text {low}}^{\text {new }}}-\frac{1}{\upsilon_{\mathrm{g}, \text {low}}^{\gamma}}\right) {\rm d} l
	 -\int \left(\frac{1}{\upsilon_{\mathrm{g}, \text {high}}^{\text {new}}}-\frac{1}{\upsilon_{\mathrm{g}, \text{high}}^{\gamma}}\right) {\rm d}  l \nonumber \\
		&\simeq&  A_1 \left(\frac{1}{\nu_{\text {low}}^{2}}-\frac{1}{\nu_{\text {high}}^{2}} \right) \mathrm{DM}
		 +A_2 \left(\frac{1}{\nu_{\text {low}}^{4}}-\frac{1}{\nu_{\text {high}}^{4}}\right) \mathrm{DM},  \nonumber \\
\end{eqnarray}
where $A_1=\frac{e^{2}}{2 \pi m_{e} c}$, $A_2=\frac{e^{2} c^3 m_{\gamma}^2}{4 \pi m_e h^2}$, and $e$ and $m_e$ are the charge and mass of an electron, respectively.
Here ${\rm DM}= \int n_e {\rm d} l$ is the dispersion measure contribution from the
plasma, which is defined as the integral of the number density of electrons $n_e$ along the line of sight. 
The second term in the last row of Equation~(\ref{new relation}) is regarded as the time delay arises from the combined effects of plasma and nonzero photon mass, which has the form of $\propto \nu^{-4}$ that differs from previous results ($\propto \nu^{-2}$; \citealt{Wu2016,Shao2017}).

\begin{figure}
	\includegraphics[width=0.45\textwidth]{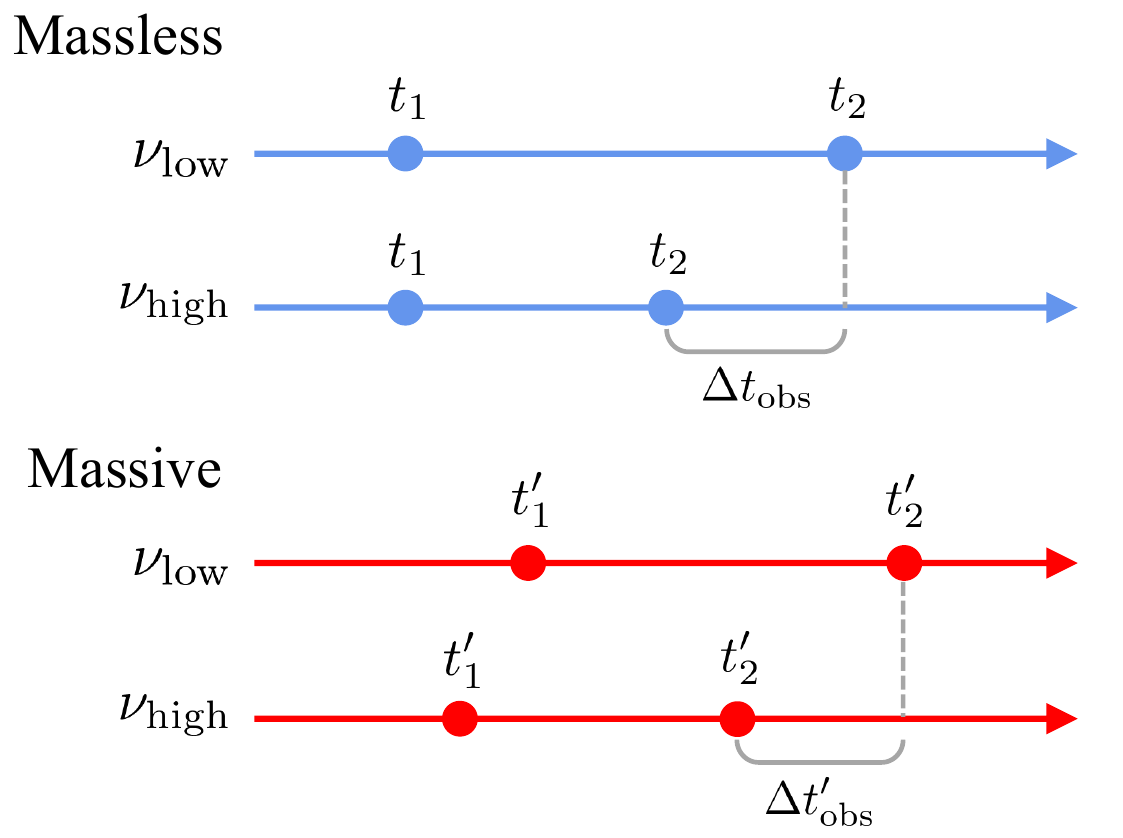}
	\caption{Timeline schematic for calculating the time delays.
		The blue and red timelines illustrate the scenarios of massless and massive photons, respectively.
		The definitions of arrival times can be referred to \autoref{Sec:comment}.
		}
	\label{Fig1}
\end{figure}

It is clear that Equation~(\ref{new relation}) (see also Equation (3) in \citealt{Wang2024}) presents a contradiction: when the electron number density $n_e$ is equal to zero, the time delay is eliminated.
This indicates that there is no dispersion for massive photons in vacuum.
\autoref{Fig1} illustrates the timelines for calculating the time delays in two scenarios: massless photons and massive photons.
In the scenario of massless photons in plasma, the observed time delay between low- and high-frequency photons is $\Delta t_{\rm obs}  = t_{\rm 2, low}-t_{\rm 2, high} = (t_{\rm 2, low}-t_{\rm 1, low} ) - ( t_{\rm 2, high}-t_{\rm 1, high} )$, where $t_{\rm 1, low} = t_{\rm 1, high}$ due to the constancy of the light speed $c$ in vacuum.
However, for massive photons, the observed time delay is $\Delta t_{\rm obs}'=   t_{\rm 2, low}' -t_{\rm 2, high}' \ne (t_{\rm 2, low}' -t_{\rm 1, low}' ) - ( t_{\rm 2, high}' -t_{\rm 1, high}' )$.
The additional subtractions of the arrival times of massive photons in vacuum (i.e., $t_{\rm 1, low}'$ and $t_{\rm 1, high}'$) in \cite{Wang2024} cause the elimination of the $\nu^{-2}$ behavior.
If we use the correct formula, then the first-order delay time term would exhibit a frequency dependence of $\nu^{-2}$, i.e.,
\begin{eqnarray}\label{correct relation}
	\Delta t_{\mathrm{obs}}^{\prime} 
		&=& t_{\rm 2, low}' - t_{\rm 2, high}' 
        = \int \left(\frac{1}{\upsilon_{\mathrm{g}, \text {low}}^{\text {new }}}-\frac{1}{\upsilon_{\mathrm{g}, \text {high}}^{\text {new}}}\right) {\rm d} l \nonumber \\
		&\simeq&  A_1 \left(\frac{1}{\nu_{\text {low}}^{2}}-\frac{1}{\nu_{\text {high}}^{2}} \right) \left(\mathrm{DM}+\mathrm{DM}_{\gamma} \right) ,
\end{eqnarray}
where the group velocity $\upsilon_{\rm g}^{\rm new}$ is expressed as
\begin{eqnarray}\label{new group velocity}
	\upsilon_{\mathrm{g}}^{\text {new }}
    &=& c  \left( 1 - \frac{m_{\gamma}^2c^4}{h^2\nu^2}\right)^{1/2} \left( 1- \frac{\nu_p^2}{\nu^2} \right)^{1/2} \nonumber \\
    &\simeq& c \left( 1 - \frac{1}{2} \frac{\nu_p^2}{\nu^2} -
    \frac{1}{2} \frac{m_{\gamma}^2c^4}{h^2\nu^2} \right) ,
\end{eqnarray}
where $\nu_p = (n_e e^2/\pi m_e)^{1/2}$ is the plasma frequency. Analogous to ${\rm DM}$, the 
``effective dispersion measure'' ($\mathrm{DM}_{\gamma}$) arising from massive photons is defined as 
$\mathrm{DM}_{\gamma} \equiv \frac{\pi m_e c^5 m_{\gamma}^2}{h^2 e^2} \int {\rm d} l$ \citep{Shao2017}. 
Equation~(\ref{correct relation}) shows that the classical dispersion relation ($\propto \nu^{-2}$) 
for massive photons propagating in plasma still holds.

\section{The Dispersion Relation Derived from Proca Equations}
\label{Sec:dispersion}

The classical Maxwell equations are founded upon the zero-mass hypothesis of the photon.
\cite{Proca1936} first considered the addition of a mass term and modified the Maxwell equations, thereby establishing the Proca equations.
The Proca Lagrangian is given as \citep{jackson2021}
\begin{equation}
	\mathcal{L}_{\rm Proca}=-\frac{1}{16\pi}F_{\alpha\beta}F^{\alpha\beta}-\frac{1}{c}J_{\alpha}A^{\alpha} + \frac{\mu^2}{8\pi}A_{\alpha}A^{\alpha}.
\end{equation}
In this Lagrangian, $F_{\alpha\beta}$ is the electromagnetic tensor,
$A_{\alpha}=(\phi,\, \mathbf{A})$ is the 4-vector potential, $J_{\alpha}=(c n_e e,\, \mathbf{J})$ is the 4-vector current, and $\mu=m_{\gamma}c/\hbar$ is the parameter related to the photon mass.
According to the Proca Lagrangian, the equation of motion can be derived as $\partial^{\beta} F_{\beta \alpha} + \mu^2 A_{\alpha} = 4 \pi J_{\alpha} /c$.
Combined with the Bianchi identity $\partial_{(\mu} F_{\alpha \beta)}=0$, the vector form of the Proca equations can be written as 
\begin{eqnarray}
		&\nabla& \cdot \mathbf{E} =4\pi n_e e - \mu^2 \phi , \label{Proca1} \\ 
		&\nabla& \times  \mathbf{E} = -  \frac{1}{c}  \frac{\partial \mathbf{B}}{\partial t } , \label{Proca2} \\  
		&\nabla& \cdot \mathbf{B} = 0, \\  
		&\nabla&  \times  \mathbf{B} = \frac{4 \pi}{c} \mathbf{J} + \frac{1}{c} \frac{\partial \mathbf{E}}{\partial t}   -  \mu^2 \mathbf{A}. \label{Proca4}
\end{eqnarray}

Once we establish the Proca equations, the dispersion relation of massive photons traveling in plasma can be derived.
We first separate the variables into two parts: an equilibrium part indicated by the subscript 0, and a perturbation part indicated by the subscript 1.
In a plasma with $\mathbf{B}_0=\mathbf{J}_0=\mathbf{E}_0=\mathbf{A}_0=0$, the linear approximation allows Equation~(\ref{Proca4}) to be converted to
\begin{equation}
	c\nabla \times  \mathbf{B_1} = 4 \pi \mathbf{J_1} + \frac{\partial \mathbf{E_1}}{\partial t}   - c\mu^2 \mathbf{A_1} .
\end{equation}
By taking the time derivative of this equation, we obtain
\begin{equation}\label{Eq8}
  c \nabla \times \frac{\partial  \mathbf{B_1} }{\partial t}= 4 \pi \frac{\partial  \mathbf{J_1}}{\partial t} + \frac{\partial^2 \mathbf{E_1}}{\partial t^2}   - c \mu^2  \frac{\partial\mathbf{A_1}}{\partial t} .
\end{equation}
The subsequent steps are to transform each term into the function of $\mathbf{E_1}$.
Taking the curl of Equation~(\ref{Proca2}), we have
\begin{eqnarray}\label{curlE}
\nabla \times \frac{\partial \mathbf{B_1}}{\partial t } & = & 
- c \nabla \times ( \nabla \times  \mathbf{E_1} ) 
= c \nabla^2 \mathbf{E_1} -  c \nabla ( \nabla  \cdot  \mathbf{E_1} ) \nonumber \\
& = & c \nabla^2 \mathbf{E_1} + c  \mu^2 \nabla \phi_1 ,
\end{eqnarray}
where the condition $\nabla  \cdot  \mathbf{E_1}= - \mu^2 \phi_1 $ from Equation~(\ref{Proca1}) is used.
Since the current is generated by electron motion, by combining the expressions $\mathbf{J_1}=n_e e\mathbf{v_1}$ and $m_e\partial \mathbf{v_1} / \partial t = -e \mathbf{E_1}$, the time derivative of the current can be expressed as
\begin{equation}\label{current}
 \frac{\partial  \mathbf{J_1}}{\partial t}  = - \frac{n_e e^2}{m_e} \mathbf{E_1} .
\end{equation}
We also need the relation between potentials ($\phi_1,\, \mathbf{A_1}$) and $\mathbf{E_1}$:
\begin{equation}\label{vector}
 \frac{\partial  \mathbf{A_1} }{\partial t} = - c \mathbf{E_1} - c \nabla \phi_1,
\end{equation}
which can be derived from the time derivative of $\mathbf{B_1} = \nabla \times \mathbf{A_1}$ and Equation~(\ref{Proca2}).
We assumed that the perturbation parts oscillate sinusoidally, i.e., $\propto \exp{[i(kx-\omega t)]}$, where $k$ and $\omega$ are the wave vector and frequency, respectively.
Hence, the time derivative and space gradient can be replaced as $\partial / \partial t \rightarrow -i \omega$ and $\nabla \rightarrow i k$.
Combining Equations~(\ref{curlE})--(\ref{vector}), Equation~(\ref{Eq8}) can be rewritten as
\begin{equation}
\left( \omega^2-c^2k^2-\omega_p^2-c^2\mu^2 \right) \mathbf{E_1} =0 ,
\end{equation}
where $\omega_p = 2 \pi \nu_p = (4 \pi n_e e^2/m_e)^{1/2}$ is the plasma frequency.
Requiring the terms in the bracket being equal to zero, one has
\begin{equation}\label{dispersion}
\omega^2 = c^2k^2+\omega_p^2+ c^2\mu^2.
\end{equation}
This is the dispersion relation of massive photons propagating in plasma.
Furthermore, the group velocity of massive photons can be calculated by
\begin{equation}\label{velocity}
\upsilon_{\rm g} = \frac{\partial \omega}{\partial k} 
= c \left(1-  \frac{ \omega_p^2+ c^2\mu^2}{ \omega^2} \right)^{1/2}   
\simeq   c \left(1-  \frac{ 1}{2} \frac{ \omega_p^2 }{ \omega^2} -  \frac{ 1}{2}\frac{ c^2\mu^2}{ \omega^2} \right) .
\end{equation}
One can see from Equation~(\ref{velocity}) that the velocity modifications from the plasma and nonzero photon mass effects are independent of each other, with the first-order expansion terms being $\propto \nu^{-2}$.
This is the natural consequence based on the Proca theory.
Therefore, the classical dispersion relation that is adopted for photon mass limits remains valid.

\begin{acknowledgments}
This work is supported by the NSFC (grant Nos. 12422307 and 12373053).
\end{acknowledgments}

\bibliographystyle{aasjournal}
\bibliography{ref}{}

\end{document}